\documentclass[preprint,12pt]{elsarticle}

\usepackage{amssymb}
\usepackage{amsmath}
\usepackage{relsize}

\usepackage{graphicx}

\journal{Physica A}

\begin{document}

\begin{frontmatter}


\title{Size scaling of failure strength at high disorder}

\author[inst1]{Zsuzsa Danku}

\affiliation[inst1]{organization={Department of Theoretical Physics, Doctoral School of Physics, Faculty of Science and Technology, University of Debrecen},
            addressline={P.O.\ Box 400}, 
            city={Debrecen},
            postcode={H-4002}, 
            country={Hungary}}

\author[inst1]{Gerg\H o P\'al}

\author[inst1,inst2]{Ferenc Kun\corref{cor1}}

\cortext[cor1]{Correspondence: ferenc.kun@science.unideb.hu (F.K.)}

\affiliation[inst2]{organization={Institute for Nuclear Research (Atomki)},
            addressline={P.O. Box 51}, 
            city={Debrecen},
            postcode={H-4001}, 
            country={Hungary}}

\begin{abstract}
We investigate how the macroscopic response and the size scaling of the ultimate strength of materials change when their local strength is sampled from a fat-tailed distribution and the degree of disorder is varied in a broad range. Using equal and localized load sharing in a fiber bundle model, we demonstrate that a transition occurs from a perfectly brittle to a quasi-brittle behaviour as the amount of disorder is gradually increased. When the load sharing is localized the high load concentration around failed regions make the system more prone to failure so that a higher degree of disorder is required for stabilization. Increasing the system size at a fixed degree of disorder an astonishing size effect is obtained: at small sizes the ultimate strength of the system increases with its size, the usual decreasing behaviour sets on only beyond a characteristic system size. The increasing regime of the size effect prevails even for localized load sharing, however, above the characteristic system size the load concentration results in a substantial strength reduction compared to equal load sharing. We show that an adequate explanation of the results can be obtained based on the extreme order statistics of fibers' strength. 
\end{abstract}

\begin{graphicalabstract}
\end{graphicalabstract}

\begin{highlights}
\item The strength of strongly disordered materials increases with the system size.
\item The usual decreasing size effect restores above a characteristic system size.
\item Stress localization leads to strength reduction above the characteristic size.
\item This size effect is controlled by the extreme order statistics of local strength.
\end{highlights}


\end{frontmatter}


\section{Introduction}
Disorder is a substantial feature of natural materials and of most of the artificially made ones. Depending on the relevant length scale, disorder can occur is various forms from the dislocations of crystalline structures on the micro-scale, to cracks, flaws, or grain boundaries at the meso-scale \cite{herrmann_statistical_1990,ciliberto_effect_2001,santucci_sub-critical_2004}. 
The presence of disorder generates weak spots in the material where cracks first nucleate when the material is subject to a slowly increasing external load \cite{alava_statistical_2006,alava_role_2008}. However, these cracks can get arrested when they penetrate into a locally stronger region and remain so until a load increment reactivates them \cite{herrmann_statistical_1990,alava_role_2008}. As a consequence, the presence of disorder, on the one hand, makes materials weaker, on the other hand, it results in an intermittent fracture process, where a large amount of damage accumulates in the sample through stable cracking, and ultimate failure occurs when merging cracks form a  macro-crack spanning the sample \cite{,petri_experimental_1994,nataf_avalanches_2014}. The nucleation and propagation of cracks generates elastic waves which can be recorded in the form of acoustic noise \cite{salje_aplmat_2023,salje_crackling_2014,vives_coal_burst_2019}. The acoustic emission technique provides a valuable insight into the intermittent dynamics of the fracture of disordered materials on the micro-scale. Cracking events can be considered as precursors of the ultimate failure of the system making it a great challenge to exploit them to forecast the imminent final collapse \cite{salje_main_minecollapse_2017,main_limits_2013,vasseur_scirep_2015,kadar_record_physa2022}.
On the macro-scale disorder has the consequence that the ultimate fracture strength exhibits sample-to-sample fluctuations, and the average strength is a decreasing function of the sample size \cite{curtin_size_1998,alava_size_2009,dill-langer_size_2003,bazant_fracture_1997}. This statistical size effect sets limits on the applicability of materials in constructions and has to be taken into account in engineering design \cite{bazant_le_2017}.

The fiber bundle model (FBM) is one of the most important modelling approaches to the fracture of heterogeneous materials. It spite of its simplicity, it grasps the essential microscopic mechanisms of fracture 
and provides an efficient way for the representation of materials' disorder \cite{hansen2015fiber,kun_extensions_2006,hidalgo_avalanche_2009}. In the simplest setup of FBMs, the sample is discretized in the form of a bundle of parallel fibers which are assumed to have a linearly elastic behaviour with the same stiffness but with a random strength. When a fiber fails, its load is redistributed among the surviving fibers. For load sharing two limiting cases have been intensively investigated during the past decades: in case of equal load sharing (ELS), all fibers receive the same load increment from the broken one resulting in a homogeneous stress field during the entire fracture process \cite{hansen2015fiber}. In the opposite limit of localized load sharing (LLS), fibers are arranged on a regular lattice and only the intact nearest neighbor fibers share the load of the broken one \cite{kadar_record_physa2022,tallakstad_local_2011,biswas_lls_2017,kun_damage_2000-1}. LLS gives rise to stress concentration around failed regions making the bundle more brittle than in the case of ELS. FBMs proved to be successful in reproducing several qualitative features of the fracture of heterogeneous materials. For instance, in FBMs the disorder implemented in the breaking thresholds of fibers has the consequence that fibers break in bursts analogous to the acoustic outbreaks of real experiments \cite{hansen2015fiber}. The macro-scale consequence of the threshold disorder is that the strength of the bundle fluctuates, furthermore, the average strength decreases with the number of fibers \cite{hansen2015fiber,smith_probability_1980,harlow_chain--bundles_1978}. Of course, the quantitative details of the behaviour of FBMs depend on the precise range of load sharing and on the degree of disorder. However, theoretical studies have revealed that for a moderate amount of disorder (e.g.\ uniform, exponential, Weibull, and Gaussian distribution of failure thresholds) under equal load sharing conditions FBMs exhibit a high degree of universality, i.e.\ the size distribution of bursts has a power law functional form with a universal exponent \cite{hansen2015fiber, kun_extensions_2006,hidalgo_avalanche_2009}, and the ultimate strength of the bundle rapidly converges to a finite value when increasing the size of the bundle \cite{smith_probability_1980,harlow_chain--bundles_1978,smith_asymptotic_1982}. The ELS size scaling is again described by a universal exponent \cite{hansen2015fiber,smith_probability_1980,harlow_chain--bundles_1978,smith_asymptotic_1982}. In case of LLS, computer simulations revealed a non-universal behaviour of the avalanche size distribution, a high degree of brittleness and a rapid decrease of the ultimate strength of the system with increasing number of fibers \cite{hidalgo_fracture_2002,PhysRevE.87.042816}.

In this paper we focus on the statistical size effect of materials' strength and study in the fiber bundle model how the interplay of stress heterogeneity caused by the localized load sharing and of the amount of disorder change the ELS behaviour. We consider a fiber bundle model with a slowly decaying fat-tailed distribution of fibers' strength. This allows us to control the degree of disorder of the bundle between the extremes of zero and infinity. We demonstrate that depending on the amount of strength disorder the macroscopic response of the system has two phases, i.e.\ perfectly brittle and quasi-brittle, where failure is triggered by the first fiber breaking, and it is preceded by an accumulation of damage, respectively. Our calculations revealed that in the case of localized load sharing a higher amount of disorder is required to stabilize the system than under ELS conditions. Increasing the number of fibers, our system exhibits an astonishing size effect: for small system sizes the bundle strength increases with the system size and the usual decreasing behaviour of the statistical size effect gets restored only above a characteristic system size. As a remarkable outcome, we demonstrate that the increasing bundle strength prevails even for localized load sharing, the ELS and LLS size effects differ only in the decreasing regime in such a way that load localization results in a considerable strength reduction.

\section{Materials and Methods}
Our study is based on the fiber bundle model where materials disorder is represented by a fat-tailed threshold distribution of fibers considering both equal and localized load sharing after local breaking events. We briefly summarize the main components of the model construction. The model has been used to study the statistics of breaking avalanches and their dependence on the system size \cite{kadar_kun_pre2019,kadar_record_2020}, furthermore, the evolution of the macroscopic strength of the bundle with the system size in the mean field limit of FBMs \cite{kadar_pre_2017}.

\subsection{Fiber bundle model with controllable disorder}
In the model the material is divided into unit elements in the form of $N$ parallel fibers arranged on a regular square lattice of side length $L$, where $N=L^2$ holds. Fibers exhibit a linearly elastic behavior with a fixed Young's modulus $E=1$, but each fiber has a unique failure threshold $\sigma_{th}=E\varepsilon_{th}$ limiting the load $\sigma$ they can hold. To capture the disorder of heterogeneous materials, the breaking thresholds are specified as the threshold strain $\varepsilon^{i}_{th} , i=1, \ldots , N$ of breaking sampled from a probability distribution $p(\varepsilon_{th})$. For $p(\varepsilon_{th})$ a fat-tailed distribution is used with a power law decay
\begin{equation}
p(\varepsilon_{th})=\begin{cases}
0, & \varepsilon_{th} < \varepsilon_{min},\\
A\varepsilon_{th}^{-(1+\mu)}, & \varepsilon_{min} \leq \varepsilon_{th} \leq \varepsilon_{max},\\
0, & \varepsilon_{max} < \varepsilon_{th},
 \end{cases}
 \label{eq:density}
\end{equation}
where the lower bound of strength is fixed to $\varepsilon_{min} = 1$, while the upper bound $\varepsilon_{max}$ can take values in the range $\varepsilon_{min} \leq \varepsilon_{max} \leq \infty$. The exponent $\mu$ controls the tail, i.e.\ the rate of decrease of the distribution. The value of $\mu$ is selected from the interval $0 \leq \mu < 1$ because for an infinite upper cutoff $\varepsilon_{max}$ in this $\mu$ range the disorder is so high that no finite average strength exists. The motivation of the choice of the specific form Eq.\ (\ref{eq:density}) of the distribution $p(\varepsilon_{th})$ is that varying the upper cutoff $\varepsilon_{max}$ of strength values and the exponent $\mu$ the amount of disorder can be varied between the limits of zero and infinity. 
\begin{figure}[h!]
\begin{center}
\includegraphics[bbllx=-1,bblly=418,bburx=455,bbury=792,scale=0.55]{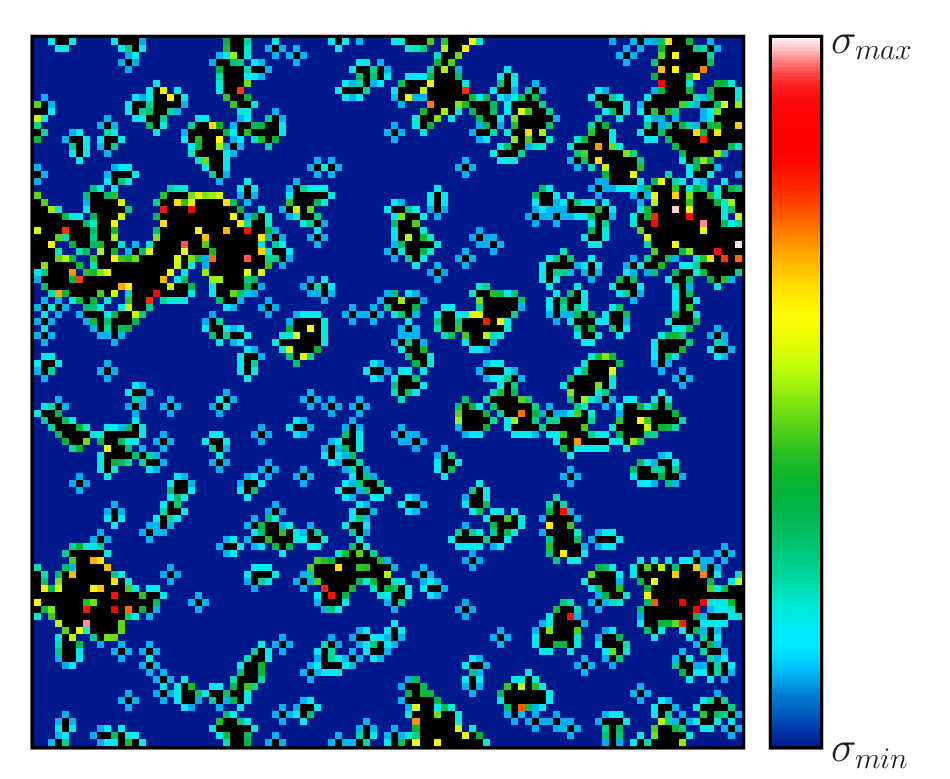}
  \caption{Fiber bundle model on a square lattice. The fibers are oriented perpendicular to the plane of the lattice. The bundle is loaded parallel to the fibers' direction. In case of ELS all intact fibers keep the same load at any stage of the loading process, while for LLS load concentration occurs along the perimeter of broken clusters. An intermediate state of an LLS bundle is presented where fibers are colored according to their load $\sigma$ in such a way that black indicates the broken fibers.
   \label{fig:demo}}
\end{center}
\end{figure}
With the normalized density function $p(\varepsilon_{th})$ the cumulative distribution $P(\varepsilon_{th})$ can be obtained as
\begin{equation}
\mathlarger{
P(\varepsilon_{th})=\begin{cases}
0, & \varepsilon_{th} < \varepsilon_{min},\\
\frac{\varepsilon_{th}^{-\mu}-\varepsilon_{min}^{-\mu}}{\varepsilon_{max}^{-\mu}-\varepsilon_{min}^{-\mu}}, & \varepsilon_{min} \leq \varepsilon_{th} \leq \varepsilon_{max},\\
1, & \varepsilon_{max} < \varepsilon_{th}.
 \end{cases}}
 \label{cumulative}
\end{equation}

The bundle is subject to a quasi-statically increasing external load $\sigma$ in such a way that $\sigma$ is increased to provoke the breaking of a single fiber. The load dropped by the broken fiber has to be overtaken by the remaining intact ones. To study how the interplay of materials' disorder and the stress field emerging during the fracture process affects the macroscopic response and size scaling of fracture strength, here we consider two limiting cases of load sharing: In case of equal load sharing (ELS) all the intact fibers receive the same load increment so that no stress inhomogeneities occur in the system. In the opposite limit of localized load sharing (LLS) the excess load is again  equally redistributed but only over the intact nearest neighbors of the broken fiber on the square lattice. As a consequence, broken fibers form spatially connected clusters and a strong load concentration emerges on the intact fibers along the cluster perimeters. For illustration see Fig.\ \ref{fig:demo}, where intact fibers are colored according to their load at an intermediate stage of the failure process. In both cases of load sharing the load increments received may induce additional breakings which are again followed by load redistribution and can trigger further breakings. Through such breaking-load redistribution steps an entire avalanche of breakings can emerge after the externally induced breaking of a single fiber. Under stress controlled loading ultimate failure of the system occurs at a critical load where a catastrophic avalanche is triggered destoying all the remaining intact fibers.

In the equal load sharing limit the macroscopic response of the system can be obtained from the general form $\sigma(\varepsilon)=E\varepsilon[1-P(E\varepsilon)]$, where the term $1-P(E\varepsilon)$ provides the fraction of intact fibers which all keep the same load $E\varepsilon$ at the strain $\varepsilon$ \cite{hansen2015fiber}. After substituting $P(\varepsilon_{th})$ from Eq.(\ref{cumulative}) and setting the Young modulus to $E=1$ the constitutive equation can be cast into the form
\begin{equation}
\mathlarger{
\sigma(\varepsilon)=\begin{cases}
\varepsilon, & 0 \leq \varepsilon < \varepsilon_{min},\\
\frac{\varepsilon(\varepsilon^{-\mu}-\varepsilon_{max}^{-\mu})}{\varepsilon_{min}^{-\mu}-\varepsilon_{max}^{-\mu}}, & \varepsilon_{min} \leq \varepsilon \leq \varepsilon_{max},\\
0, & \varepsilon_{max} < \varepsilon.
 \end{cases}
 }
 \label{eq:sigeps}
\end{equation}
The fat tailed strength distribution has a substantial effect on the fracture process both on the macro- and micro-scales. Here we explore how the macroscopic response and the size scaling of the ultimate strength of the bundle changes when the amount of disorder is varied on the $\mu-\varepsilon_{max}$ plane considering both equal and localized load sharing after the breaking of fibers.

\section{Results}

\subsection{Disorder driven brittle to quasi-brittle transition}
\begin{figure}[h!]
\begin{center}
\includegraphics[bbllx=0,bblly=0,bburx=388,bbury=343,scale=0.8]{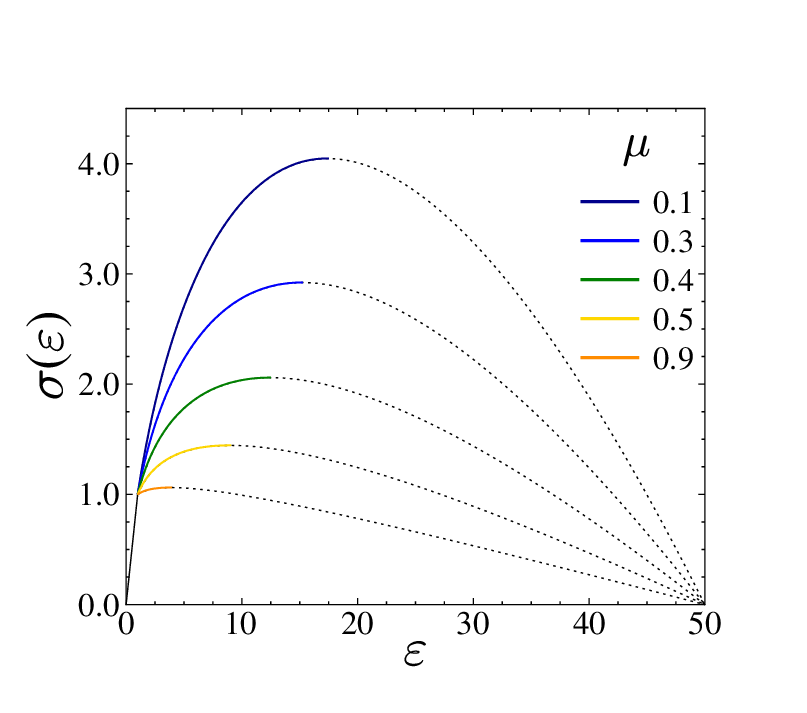}
  \caption{Constitutive curves of the model for several values of the exponent $\mu$ at the same upper cutoff $\varepsilon_{max}=50$ of fibers' strength using equal load sharing after fiber breaking. Reducing disorder by increasing the exponent $\mu$ the response of the system becomes more-and-more brittle.
   \label{fig:phasediag}}
\end{center}
\end{figure}
To understand how the amount of strength disorder affects the macroscopic response of the bundle, Fig.\ \ref{fig:phasediag} presents stress-strain curves $\sigma(\varepsilon)$ Eq.\ (\ref{eq:sigeps}) of equal load sharing bundles for different values of the exponent $\mu$ while the upper cutoff $\varepsilon_{max}$ is fixed. Since the breaking thresholds have a finite lower bound $\varepsilon_{min}$, no failure can occur at small deformation $\varepsilon < \varepsilon_{min}$, so that the system has a perfectly linearly elastic response in this range. As the value of $\varepsilon_{min}$ is surpassed during the loading process, fiber breaking sets on and the behavior of the bundle becomes non-linear. It can be observed that under ELS conditions  the constitutive curves $\sigma(\varepsilon)$ have a maximum whose value $\sigma_c$ and position $\varepsilon_c$ define the critical stress and strain where ultimate failure occurs under stress controlled loading. Both strength values $\sigma_c$ and $\varepsilon_c$ depend on the degree of disorder, i.e.\ on the parameters $\varepsilon_{max}$ and $\mu$, which can be obtained analytically from Eq.\ (\ref{eq:sigeps}) as
\begin{equation}
\varepsilon_c=\varepsilon_{max}(1-\mu)^{1/\mu},
\label{eps_c}
\end{equation}
and
\begin{equation}
\sigma_c=\frac{\mu(1-\mu)^{1/\mu-1}\varepsilon_{max}^{1-\mu}}{\varepsilon_{min}^{-\mu}-\varepsilon_{max}^{-\mu}}.
\label{sig_c}
\end{equation}
Slowly increasing the external load on the bundle at the maximum of the $\sigma(\varepsilon)$ curve a catastrophic avalanche of breaking is initiated which destroys all the intact fibers. In case of localized load sharing the constitutive behaviour $\sigma(\varepsilon)$ of FBMs follow the same curve as in ELS, however, usually with a lower ultimate strength, i.e.\ the $\sigma(\varepsilon)$ curves stop at lower loads implying a more brittle response \cite{raischel_local_2006,PhysRevE.87.042816}.

\begin{figure}[h!]
\begin{center}
\includegraphics[bbllx=0,bblly=0,bburx=390,bbury=330,scale=0.8]{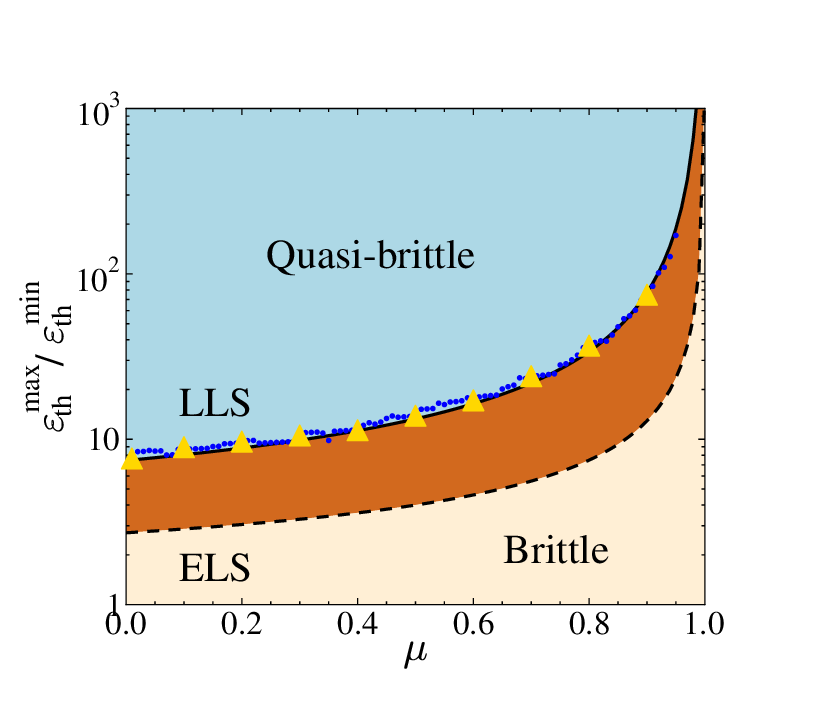}
  \caption{Phase diagram of the bundle with power law distributed breaking thresholds for both equal (dashed line, ELS) and localized load sharing (dots with a fitted continuous line, LLS). The phase boundary separates the perfectly brittle behaviour where the first fiber breaking triggers catastrophic collapse and the quasi-brittle phase where global failure is preceded by stable damaging. At each value of the disorder exponent $\mu$ the LLS curve lies above the ELS phase boundary so that the area between the two curves is brittle for LLS but is already quasi-brittle under ELS conditions. LLS calculations represented by dots were performed with the system size $L=1001$, while the triangles were obtained with a larger system $L=2001$.
   \label{fig:phasediag_LLS}}
\end{center}
\end{figure}

Increasing the value of the exponent $\mu$ or decreasing the upper bound $\varepsilon_{max}$, the threshold disorder decreases, until already the first fiber breaking is able to trigger the macroscopic failure of the whole system. In Fig.\ \ref{fig:phasediag} this behaviour can be observed in such a way that as the value of $\mu$ increases at a fixed cutoff $\varepsilon_{max}$, both the height $\sigma_c$ and the position $\varepsilon_c$ of the maximum decrease and approach the end point of the linear regime of the $\sigma(\varepsilon)$ curve. It follows that reducing the amount of disorder in the bundle the macroscopic response exhibits a transition from quasi-brittle where a macroscopic failure is preceded by a large amount of damage to perfectly brittle where failure occurs abruptly at the instant of the first fiber breaking.
At each value of the disorder exponent $\mu$ there exists a critical upper bound $\varepsilon^c_{max}$ of strength values which separates the two regimes of qualitatively different behaviours. In the limit of equal load sharing the phase boundary can be obtained analytically from Eqs. (\ref{eps_c}) and (\ref{sig_c}) as 
\begin{equation}
\varepsilon^c_{max}=\frac{\varepsilon_{min}}{(1-\mu)^{1/\mu}}.
\label{eq:elsphasebound}
\end{equation}
It can be inferred that as the exponent approaches one from below $\mu \to 1$, the critical value of the upper bound diverges $\varepsilon^c_{max} \to \infty$, so in the regime $\mu \geq 1$ the bundle always has a brittle response independent of the value of $\varepsilon_{max}$. These findings are summarized in Fig.\ \ref{fig:phasediag_LLS} which shows the phase boundary of the system on the $\mu-\varepsilon_{max}$ plane. 

It is a question of fundamental importance how the range of load redistribution affects the macroscopic response of the system and its dependence on the degree of disorder. When the load is redistributed locally on the nearest neighbors of broken fibers a high stress concentration occurs, which makes the intact fibers along the perimeter of broken clusters more prone to failure (see Fig.\ \ref{fig:demo}). As a consequence, the failure process of fibers will be determined by the competition of the inhomogeneous stress field and the disordered fiber strength, where strong fibers may break before weak ones if they received a high load increment from their broken neighbors \cite{hansen_lls_dimension_2015,kjellstadli_shielding_2019,PhysRevE.87.042816}. To determine the  boundary between the phases of brittle and quasi-brittle responses under LLS conditions we performed a large amount of computer simulations with the system size $L=1001$ collecting data of $K=10^4$ samples at each parameter set considered. For a given value of the exponent $\mu$, calculations started at an upper bound $\varepsilon_{max}$ where the system has a perfectly brittle behaviour, i.e.\ where all the $K$ samples suffer catastrophic failure right at the breaking of the weakest fiber. The value of $\varepsilon_{max}$ was gradually increased in small steps and the number of samples was monitored which suffered catastrophic collapse at the first fiber breaking. The LLS critical value $\varepsilon_{max}^c$ was identified as the upper bound of fibers' strength where the number of collapsing systems first drops to zero. The numerical results are presented in Fig.\ \ref{fig:phasediag_LLS} along with the mean field (ELS) phase boundary. Motivated by the functional form of the corresponding ELS curve Eq.\ (\ref{eq:elsphasebound}),  we fitted the results with the expression
\begin{equation}
 f(\mu) = \frac{a}{(1-\mu)^{b/\mu}},
\end{equation}
where best fit was obtained with the parameter values $a=1.66$ and $b=1.5$.
It can be observed in Fig.\ \ref{fig:phasediag_LLS} that the LLS phase boundary lies everywhere above the ELS curve, which indicates that in the presence of stress concentrations a higher amount of strength disorder is needed to stabilize the system. Numerically this is indicated by the result that both parameters $a$ and $b$ are greater than their ELS counterparts, $\varepsilon_{min}$ and 1, respectively. Note that as the exponent $\mu$ increases towards 1, the difference between LLS and ELS gradually disappears since the system becomes totally brittle, where the range of load sharing does not play any role. 

To test how the system size affects the phase boundary in the LLS case, simulations were repeated at a significantly larger lattice size $L=2001$, the results of which are highlighted by the triangles in Fig.\ \ref{fig:phasediag_LLS}. It can be observed that in this range of $L$ the phase boundary can be considered to be independent of the system size. For the sake of simplicity, to characterize how far the system is from the phase boundary at a given value of the upper cutoff $\varepsilon_{max}$, we introduce the parameter $k=\varepsilon_{max}/\varepsilon_{max}^c(\mu)$, which takes values in the range $k\geq 1$ for any exponent $\mu$. 

\subsection{Size scaling of the ultimate strength of the bundle}
The disordered strength of local material elements implies that the overall strength has sample to sample fluctuations and the average strength decreases with the sample size. In the framework of fiber bundle models it has been shown analytically that in the case of equal load sharing the average bundle strength $\left<\sigma_c\right>$ and $\left<\varepsilon_c\right>$ monotonically decrease and converge towards finite asymptotic values according to power laws
\begin{align}
    \left<\sigma_c\right>(N) &= \sigma_c(\infty)+AN^{-\alpha}, \\
    \left<\varepsilon_c\right>(N) &= \varepsilon_c(\infty)+BN^{-\alpha},
    \label{eq:sizescalingels}
\end{align}
where $\sigma_c(\infty)$ and $\varepsilon_c(\infty)$ denote the asymptotic strength \cite{bazant_le_2017,hansen2015fiber,smith_probability_1980,harlow_chain--bundles_1978}. The power law exponent $\alpha$ proved to have the value $\alpha=2/3$, which is universal for the class of moderate disorder, where the probability density function of failure thresholds converges to zero sufficiently fast in the limit of large thresholds. Computer simulations have confirmed the validity of Eqs.\ (\ref{eq:sizescalingels}) for the uniform, Gaussian, and Weibull distributions \cite{hansen2015fiber}. 

\begin{figure}[h!]
\begin{center}
\includegraphics[bbllx=0,bblly=0,bburx=711,bbury=314,scale=0.5]{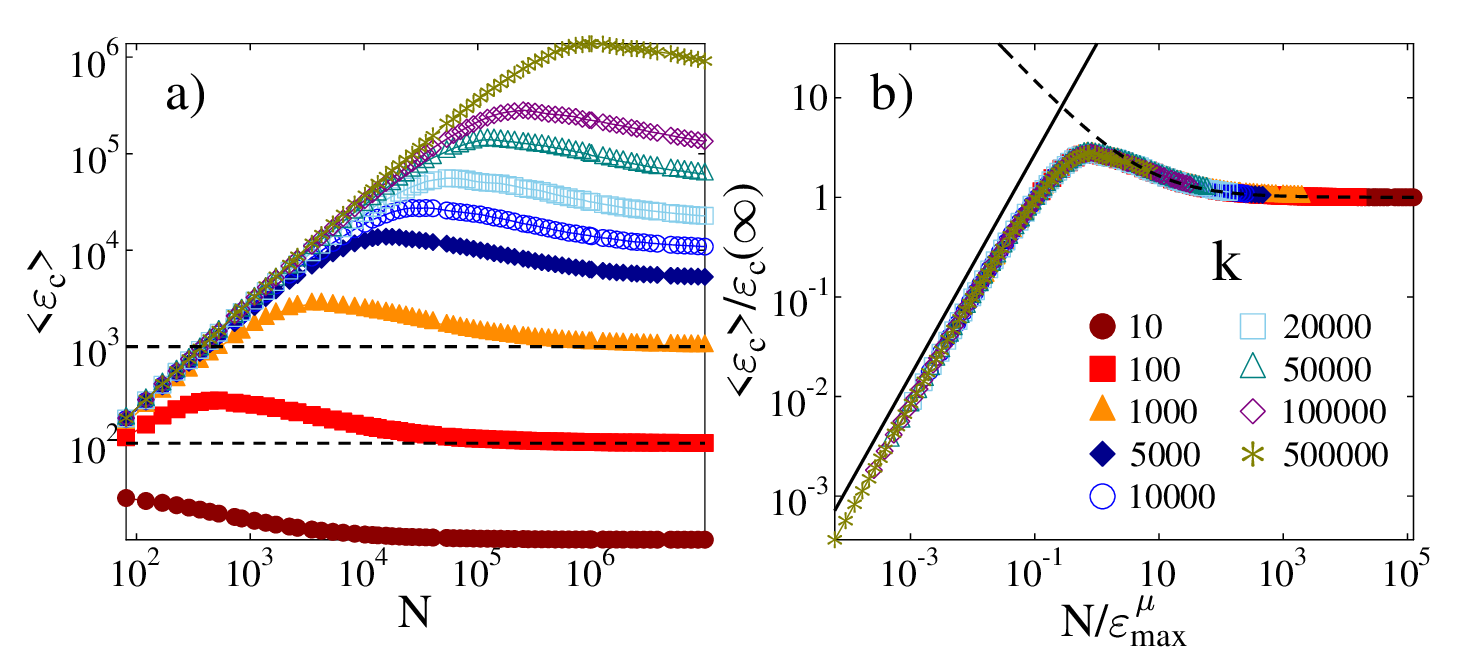}
  \caption{
   \label{fig:sizeeffectels} $(a)$ Average critical strain of failure $\left<\varepsilon_c\right>$ as a function of the number of fibers $N$ for several values of the upper cutoff of fibers' strength characterized by the parameter $k$. The value of the disorder exponent is fixed to $\mu=0.9$. The horizontals dashed lines highlight the value of the asymptotic strength for $k=100$ and $k=1000$. $(b)$ Rescaling the $\left<\varepsilon_c\right>$ curves of $(a)$ with $\varepsilon_{max}^{\mu}$ along the horizontal axis and with the corresponding asymptotic strength $\varepsilon_c(\infty)$ obtained from Eq.\ (\ref{eps_c}), curves of different disorder parameters can be collapsed on a master curve. The straight line represents the evolution of the average strength of the strongest fiber with the system size $N$ Eq.\ (\ref{eq:strength_pow}), while the dashed line indicates the ELS size scaling Eq.\ (\ref{eq:sizescalingels}). }
\end{center}
\end{figure}

Recently, we have shown that under ELS conditions the fat tailed disorder Eq.\ (\ref{eq:density}) of our bundle gives rise to a peculiar size effect: for small system sizes $N$ the strength of the bundle increases with increasing number of fibers and the usual decreasing behaviour sets on only beyond a characteristic system size $N_c$. This size effect is illustrated in Fig.\ \ref{fig:sizeeffectels}$(a)$ for $\left<\varepsilon_c\right>$ varying the upper cutoff of strength values $\varepsilon_{max}$ at a fixed disorder exponent $\mu=0.9$. It can be seen that the maximum strength and the characteristic system size $N_c$ where the increasing regime ends and the decreasing behaviour sets on strongly depend on the disorder parameters. Fig.\ \ref{fig:sizeeffectels}$(b)$ demonstrates that rescaling the $\left<\varepsilon_c\right>(N)$ curves with a proper power of the upper cutoff $\varepsilon_{max}$, results obtained at different cutoffs can be collapsed on a master curve. Our numerical analysis confirm that the scaling exponent coincides with the disorder exponent $\mu$ so that the size dependent sample strength obeys the scaling law
\begin{align}
        \left<\varepsilon_c\right>(N) &= \varepsilon_c(\infty) F(N/\varepsilon_{max}^{\mu}), \\
        \left<\sigma_c\right>(N) &= \sigma_c(\infty) G(N/\varepsilon_{max}^{\mu}),
        \label{eq:collapse}
\end{align}
where $F(x)$ and $G(x)$ denote the scaling function of the critical strain and stress, respectively.

The origin of this interesting size effect is the fat tailed nature of the strength disorder of fibers: since the disorder distribution Eq.\ (\ref{eq:density}) slowly decays it has a considerable probability even at relatively small system sizes that the bundle contains very strong fibers with breaking thresholds close to the upper bound $\varepsilon_{max}$. These fibers can be so strong that one or a few of them maybe able to keep the entire load of the system after the weaker ones have already failed. It follows from this argument that the global strength of the bundle is determined by the strongest fibers, i.e.\ by the extreme order statistics of fibers' strength. The average value of the largest element $\left<\varepsilon_{th}\right>_N$ of a sequence of $N$ random thresholds $\varepsilon_{th}^i, i=1,\ldots,N$ sampled from the same distribution independently of each other can be written in the form \cite{hansen_statistical_2000}
\begin{equation}
     \left<\varepsilon_{th}\right>_N=P^{-1}\left(1-\frac{1}{N+1}\right).
    \label{eq:extreme}
\end{equation}
Inserting the cummulative distribution of thresholds $P$ we obtain that the average strength of a bundle of $N$ fibers $\left<\varepsilon_c\right>(N)$ increases as a power law of $N$ 
\begin{equation}
\left<\varepsilon_c\right>(N)\approx \left<\varepsilon_c\right>(N) \sim N^{1/\mu}.
\label{eq:strength_pow}
\end{equation}
Beyond a certain system size due to the large number of weak fibers the strong ones cannot keep the high load accumulating as the weaker fibers break. Consequently, the usual decreasing behaviour of the ultimate strength gets restored. Figure \ref{fig:sizeeffectels}$(b)$ demonstrates that the functional form Eq.\ (\ref{eq:strength_pow}) provides a reasonable description of the increasing regime of the scaling function, while the decreasing branch is well described by the general ELS size effect Eq.\ (\ref{eq:sizescalingels}). Note that the size scaling exponent of $\left<\varepsilon_{c}\right>(N)$  in the increasing regime Eq.\ (\ref{eq:strength_pow}) depends on the degree of disorder, however, it has the universal ELS value $\alpha=2/3$ in the decreasing branch. This shows again the robustness of the ELS universality class of FBMs, which extends also to fat-tailed disorder if the system size is sufficiently large.

The transition between the increasing and decreasing regimes of the size dependent strength occurs at a characteristic value of the system size $N_c$, which must depend on the degree of disorder of fibers' strength.
Based on the above explanation in terms of extreme order statistics, we can derive an approximate analytic expression for $N_c$: the strongest fiber can dominate the macroscopic behaviour of the bundle until it can keep the entire load dropped by the $N-1$ broken ones. Let's consider a simplified bundle of size $N$ where the breaking thresholds are equal to their respective average $\left<\varepsilon_{th}\right>_i$ ($i=1,\ldots, N$) in an ordered sequence. Here $\left<\varepsilon_{th}\right>_i$ denotes the average value of the $i$th largest threshold, which can be obtained as 
\begin{align}
\mathlarger{
    \left<\varepsilon_{th}\right>_i =P^{-1}\left(\frac{i}{N+1}\right),}
\end{align}
which yields 
\begin{align}
    \left<\varepsilon_{th}\right>_i = \left[\varepsilon_{min}^{-\mu}+\frac{i}{N+1}\left(\varepsilon_{max}^{-\mu}-\varepsilon_{min}^{-\mu}\right)\right]^{-1/\mu},
\end{align}
after substituting the cumulative threshold distribution $P$ \cite{hansen_statistical_2000}. If the bundle size $N$ falls in the increasing regime in Fig.\ \ref{fig:sizeeffectels}, we can assume that the bundle can be loaded till the last failure threshold is reached.
When the slowly increasing external load reaches the failure threshold of the $(N-1)$th fiber, the total load on the system is $2\left<\varepsilon_{th}\right>_{N-1}$. To guarantee that the last fiber can keep the entire load, the condition $2\left<\varepsilon_{th}\right>_{N-1}<\left<\varepsilon_{th}\right>_{N}$ must hold, which results in a condition for the system size $N$. It follows that the value of $N$ has to be smaller than $N_c$, where the characteristic system size $N_c$ can be expressed in terms of the disorder parameters as
\begin{align}
\mathlarger{
    N_c =\frac{2^{-\mu}(\varepsilon_{max}^{-\mu}-\varepsilon_{min}^{-\mu})-\varepsilon_{min}^{-\mu}(2^{-\mu}-1)}{(2^{-\mu}-1)\varepsilon_{max}^{-\mu}}.}
    \label{eq:nc_anal}
\end{align}
Taking the limit when the upper bound of breaking thresholds is much larger than the lower one $\varepsilon_{max}\gg\varepsilon_{min}$, the asymptotics of Eq.\ (\ref{eq:nc_anal}) reads as
\begin{align}
    N_c\approx \left(\frac{\varepsilon_{max}}{\varepsilon_{min}}\right)^{\mu}\frac{2^{1-\mu}-1}{1-2^{-\mu}}.
\end{align}
Note that this asymptotic result is consistent with the behaviour of $N_c\sim \varepsilon_{max}^{\mu}$, which can be inferred from the scaling analysis Eq.\ (\ref{eq:collapse}).

\begin{figure}[h!]
\begin{center}
\includegraphics[bbllx=0,bblly=0,bburx=700,bbury=306,scale=0.54]{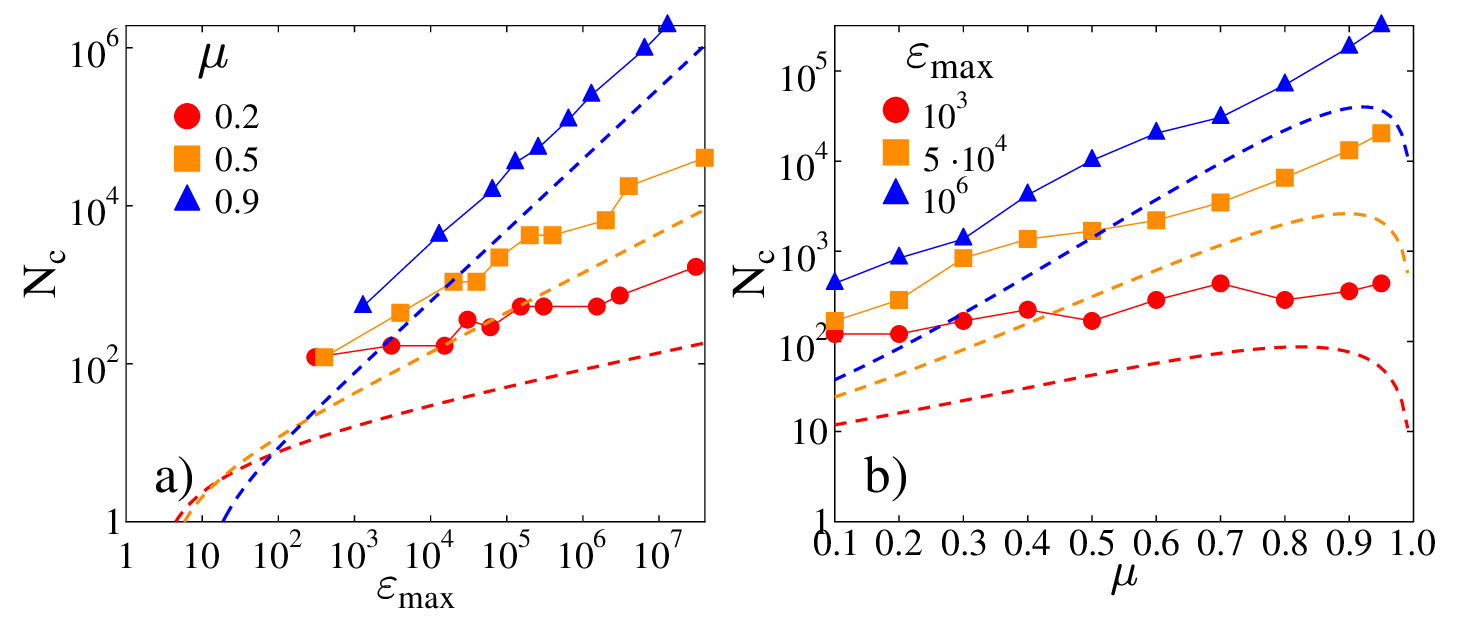}
  \caption{The critical system size $N_c$ where the ultimate strength reaches its maximum value for ELS bundles. Numerical results (symbols) are compared to the analytical predictions (dashed lines) of Eq.\ (\ref{eq:nc_anal}) as a function of the cutoff strength $\varepsilon_{max}$ at fixed disorder exponents $\mu$ $(a)$, and as a function of $\mu$ for fixed cutoffs $\varepsilon_{max}$ $(b)$.
  \label{fig:N_c}}
\end{center}
\end{figure}
For the case of ELS we also determined the value of the critical system size $N_c$ directly by computer simulations. In Figure \ref{fig:N_c} numerical results are compared to the analytical predictions of Eq.\ (\ref{eq:nc_anal}) as functions of the cutoff strength $\varepsilon_{max}$ and of the disorder exponent $\mu$, while the value of the other parameter is fixed.
It can be observed that at a fixed value of the disorder exponent $\mu$, the characteristic system size $N_c$ is an increasing function of the cutoff strength $\varepsilon_{max}$ of fibers, which is also confirmed by the numerical findings (see Fig.\ \ref{fig:N_c}$(a)$). The reason is that increasing $\varepsilon_{max}$ the strongest fibers become stronger in the bundle so that they can keep the load of a larger number of weak fibers. It is interesting to note that fixing  $\varepsilon_{max}$ the value of $N_c$ increases also with the disorder exponent $\mu$ (see Fig.\ \ref{fig:N_c}$(b)$). The result is somewhat counter intuitive since at a higher exponent $\mu$ the strength distribution decays faster reducing the fraction of strong fibers. The explanation of the increasing $N_c$ is that increasing $\mu$ the fraction of the weakest fibers also increases substantially, i.e. the ones which have a strength close to the lower bound $\varepsilon_{min}$, which can be still balanced by the fewer very strong fibers from the vicinity of the upper bound $\varepsilon_{max}$. Of course, when the exponent gets to high values close to $\mu=1$, the situation changes, i.e.\ at such low disorders the small number of strong fibers cannot keep the load dropped by the large number of weak ones anymore, hence the value of $N_c$ has a maximum at a certain value of $\mu$ beyond which it rapidly drops to zero. It can be observed in Fig.\ \ref{fig:N_c} that apart from the sharp decrease in the vicinity of $\mu=1$, the numerical and analytical results are in a reasonable agreement with each other.

\begin{figure}[h!]
\begin{center}
\includegraphics[bbllx=0,bblly=0,bburx=703,bbury=306,scale=0.55]{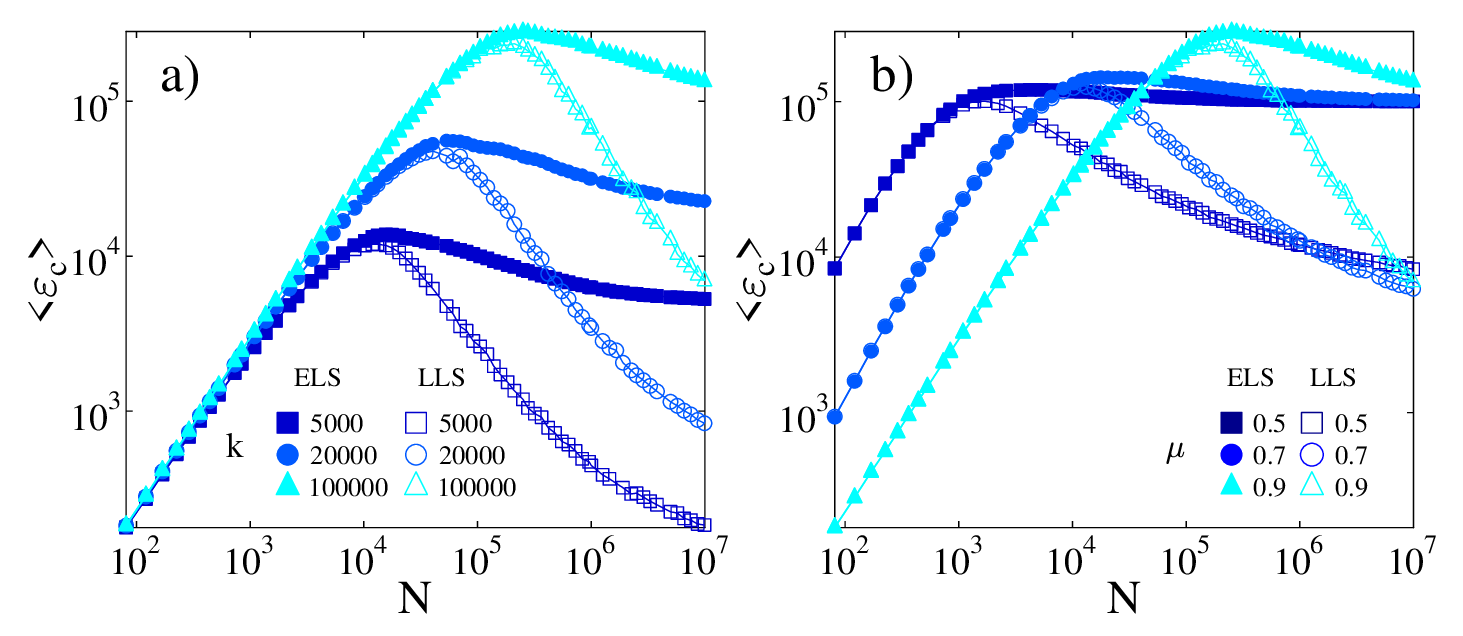}
  \caption{Comparison of the size scaling of the average critical strain $\left<\varepsilon_c\right>$ for equal and localized load sharing for three values of the cutoff strength parameter $k$ at a fixed disorder exponent $\mu=0.9$ $(a)$, and for three $\mu$ values at a fixed upper cutoff $k=100000$ $(b)$. 
   \label{fig:sizeffectlls}}
\end{center}
\end{figure}

It is a crucial question of practical importance how the range of load sharing affects the statistical size effect. We performed large scale LLS simulations on square lattices varying the lattice size from $L=9$ to $L=3151$ so that the number of fibers covered the range $N=81-9,928,801$. Figure \ref{fig:sizeffectlls} compares the evolution of the average critical strain $\left<\varepsilon_c\right>(N)$ of ELS and LLS bundles with the system size $N$ varying the upper cutoff $(a)$ and the exponent $(b)$ of the strength distribution of fibers. It is interesting to note that up to the maximum the ELS-LLS pairs of curves coincide with each other which implies that until the very strong fibers dominate the ultimate strength of the bundle the stress fluctuations due to localized load sharing cannot have a relevant effect. This means that even in the case of strongly localized load sharing the bundle strength increases with the number of fibers $N$ for bundle sizes $N<N_c$. Beyond the maximum strength, both the ELS and LLS values decrease as expected, however, deviations of the ELS and LLS curves become larger and larger with increasing number of fibers $N$ so that the LLS strength becomes significantly lower than the ELS one at the same $N$. 

\begin{figure}[h!]
\begin{center}
\includegraphics[bbllx=0,bblly=0,bburx=757,bbury=306,scale=0.5]{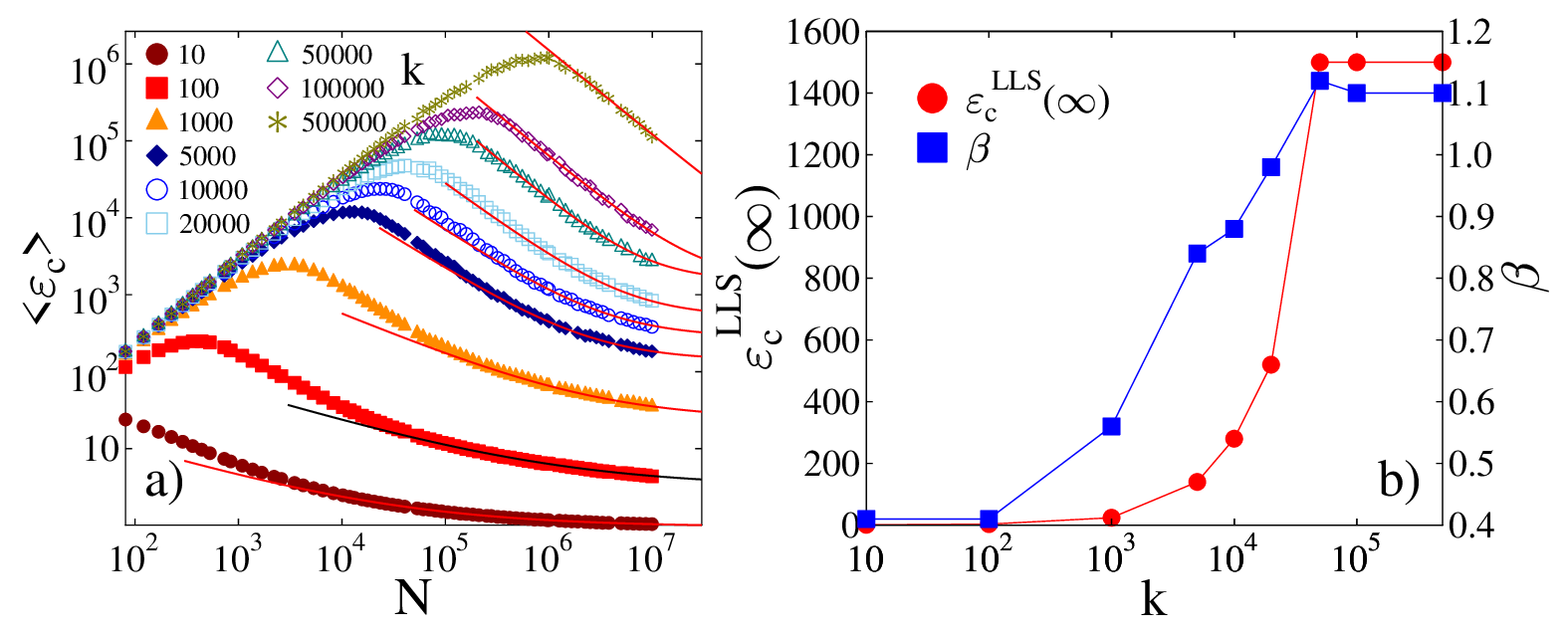}
  \caption{$(a)$ Fitting of the decreasing regime of the size scaling of the ultimate strength of LLS bundles with the functional form Eq.\ (\ref{eq:llssizescaling}) for several cutoff values $k$. $(b)$ The asymptotic strength $\varepsilon_c^{LLS}(\infty)$ and size scaling exponent $\beta$ of LLS bundles obtained by fitting as function of the cutoff parameter $k$ at a fixed value of the disorder exponent $\mu=0.9$.
   \label{fig:llsstrength}}
\end{center}
\end{figure}

Since the strength of fibers has a non-zero lower bound $\varepsilon_{min}$, the LLS bundle strength must also be bounded from below although it is not so apparent in the range of $N$ considered in Fig.\ \ref{fig:sizeffectlls}. 
To get a quantitative insight into the evolution, as an approximation, we fitted the LLS size scaling curves with the expression
\begin{equation}
    \left<\varepsilon_c\right>(N) = \varepsilon_c^{LLS}(\infty)+DN^{-\beta},
    \label{eq:llssizescaling}
\end{equation}
where $\varepsilon_c^{LLS}(\infty)$ denotes the asymptotic strength of the LLS bundle and $\beta$ is the scaling exponent. It can be observed in Fig.\ \ref{fig:llsstrength}$(a)$ that Eq.\ (\ref{eq:llssizescaling}) provides a reasonable description of the numerical results. As the amount of disorder increases with increasing upper cutoff $k$, the quality of fitting gradually improves.
The asymptotic strength $\varepsilon_c^{LLS}(\infty)$ and the size scaling exponent $\beta$ of LLS bundles obtained by fitting are presented in Fig.\ \ref{fig:llsstrength}$(b)$ as function of the cutoff strength of fibers at a fixed exponent $\mu=0.9$. The value of the size scaling exponent $\beta$ covers the range $0.4<\beta<1$ so that it can be smaller and larger than the corresponding ELS value $\alpha=2/3$. The increasing value of $\varepsilon_c^{LLS}(\infty)$ and $\beta$ implies that at higher disorder a higher asymptotic strength is obtained which is approached faster with increasing number of fibers $N$. It is interesting to note that at moderate disorder, where the threshold distribution has a fast decay towards high strength values, a logarithmic size effect has been predicted for localized load sharing fiber bundles. Both the increasing regime of size scaling at small system sizes and the power law decrease for large ones are the consequence of the slowly decaying fat-tailed strength disorder of fibers.

\section{Discussion and conclusions}
Disorder plays a crucial role in failure processes of materials. Typically disorder makes materials weaker in the sense that cracks can be initiated at weak spots earlier at lower loads, which can eventually reduce their ultimate strength. However, disorder makes it possible to arrest propagating cracks when they enter locally stronger regions resulting in stabilization. As a consequence, in the presence of a sufficiently high amount of disorder, ultimate failure occurs as the culmination of an intermittent accumulation of damage. When stress concentrations are present in the system around failed regions, a complex competition emerges between the disordered strength of material elements and the inhomogeneous stress field, where stronger regions carrying a higher load may fail before weaker ones. On the macroscale, disorder gives rise to fluctuations of the ultimate strength of materials such that the average strength tends to decrease with the system size.

Based on a fiber bundle model of heterogeneous materials, here we investigated how the macroscopic response and the ultimate strength changes when the local strength of fibers is sampled from a fat-tailed distribution. A power law distribution of breaking thresholds was considered which allowed us to control the degree of disorder by varying the power law exponent and the upper cutoff of strength values. The main goal of our study was to explore how the interplay of the inhomogeneous stress field and of the strength disorder affects the brittle and quasi-brittle nature of the macroscopic response and the size dependence of the ultimate strength of the bundle.

In the limit of a low amount of threshold disorder the macroscopic response of the system is perfectly brittle, which means that already the first fiber breaking triggers the immediate collapse of the bundle. At each disorder exponent, there exists a critical value of the upper cutoff of the strength distribution of fibers, which has to be surpassed to keep the bundle stable after the first breaking and obtain a quasi-brittle response where global failure is approached through a sequence of breaking avalanches. In the limit of equal load sharing this phase boundary can be determined analytically. Calculations showed that at higher exponents, where the strength distribution of fibers decays faster, a higher cutoff strength is required to obtain a quasi-brittle response. When the load sharing is localized, the load concentration around failed fibers makes the system more prone to failure. As a consequence, a higher amount of strength disorder is needed to stabilize the system so that the LLS phase boundary lies above the ELS one on the plane of disorder parameters. Computer simulations revealed that at high disorder exponents the difference of ELS and LLS bundles gradually disappears. 

As a very interesting feature of our system we demonstrated that the global strength of the bundle exhibits a peculiar size scaling: for small system sizes the strength of the bundle is found to increase with the number of fibers and the usual decreasing behaviour gets restored only above a characteristic system size. This unique size effect is the consequence of the fat-tailed nature of the strength distribution of fibers: at small system sizes very strong fibers with threshold values close to the cutoff strength may occur with a considerable probability. These fibers can be sufficiently strong so that even a single one may be able to keep the entire load after all the other fibers have failed. It follows that the overall strength of the system is determined by the strength of the strongest fiber, i.e. by the extreme order statistics of fibers' strength. However, there exists a characteristic system size above which the large amount of load kept by the week fibers overcomes the strength of the strongest fibers of the bundle leading to the usual decreasing behaviour of global strength. Based on the above argument we gave an analytical description of the increasing regime of the ultimate strength and derived an approximate analytic expression for the characteristic system size at which the maximum strength is obtained in terms of the two disorder parameters of the model.

Localized load sharing is known to have a strong effect on the size scaling of fracture strength typically resulting in a logarithmic decrease towards the lower bound of the strength of single fibers.
As the most important outcome we showed that the increasing strength with the system size remains valid even for localized load sharing in such a way that the size dependence of the ultimate strength of ELS and LLS bundles coincide up to the characteristic system size of maximal strength. For system sizes in the decreasing regime, LLS bundles always have a lower strength than their ELS counterparts. To give a quantitative characterization of the size scaling of LLS bundles, in the decreasing regime a power law functional form was used to fit the convergence of strength towards a finite lower bound. The numerical analysis revealed that increasing the cutoff strength of fibers at a fixed value of the disorder exponent both the asymptotic strength of the bundle and the scaling exponent increases, which implies that at higher disorder a higher strength is obtained towards which the system converges faster. Our results demonstrate that fat-tailed disorder has a strong effect on the macro-scale characteristics of fracture processes irrespective of the range of load sharing which could be exploited for materials design.

\section*{Acknowledgments}
The work was supported by the EFOP-3.6.1-16-2016-00022 project. This research was supported by the National Research, Development and Innovation Fund of Hungary, financed under the K-16 funding scheme Project no. K 119967. Project no. TKP2020-NKA-04 has been implemented with the support provided from the National Research, Development and Innovation Fund of Hungary, financed under the 2020–4.1.1-TKP2020 funding scheme. 


 \bibliographystyle{elsarticle-num} 
 \bibliography{statphys_fracture}





\end{document}